\documentclass[epj]{webofc}
\usepackage[varg]{txfonts}   
\wocname{EPJ Web of Conferences}
\woctitle{ICNFP 2016}
\begin{document}
\selectlanguage{english}
\title{Doubly hidden-charm/bottom $QQ\bar Q\bar Q$ tetraquark states}
%
%

\author{Wei Chen\inst{1}\fnsep\thanks{\bf{Speaker}} \fnsep\thanks{\email{chenwei29@mail.sysu.edu.cn}}
\and
        Hua-Xing Chen\inst{2} \and
        Xiang Liu\inst{3, 4} \and
        T. G. Steele\inst{5} \and
        Shi-Lin Zhu\inst{6, 7, 8}
}

\institute{School of Physics, Sun Yat-Sen University, Guang Zhou 510275, China
\and
           School of Physics and Beijing Key Laboratory of Advanced Nuclear Materials and Physics, Beihang University, Beijing 100191, China 
\and
           School of Physical Science and Technology, Lanzhou University, Lanzhou 730000, China
\and 
Research Center for Hadron and CSR Physics, Lanzhou University and Institute of Modern Physics of CAS, Lanzhou 730000, China
\and 
Department of Physics and Engineering Physics, University of Saskatchewan, Saskatoon, Saskatchewan, S7N 5E2, Canada
\and 
School of Physics and State Key Laboratory of Nuclear Physics and Technology, Peking University, Beijing 100871, China
\and 
Collaborative Innovation Center of Quantum Matter, Beijing 100871, China
\and
Center of High Energy Physics, Peking University, Beijing 100871, China
}

\abstract{%
We study the mass spectra for the $cc\bar c\bar c$ and $bb\bar b\bar b$ tetraquark states 
by developing a moment sum rule method. Our results show that the $bb\bar b\bar b$ tetraquarks lie 
below the threshold of $\eta_b(1S)\eta_b(1S)$. They are probably stable and very narrow. 
The masses for the doubly hidden-charm states $cc\bar c\bar c$ are higher than the spontaneous 
dissociation thresholds of two charmonium mesons. We suggest to search for such states in the 
$J/\psi J/\psi$ and $\eta_c(1S)\eta_c(1S)$ channels.
}
\maketitle
\section{Introduction}
\label{intro}
The configurations of multiquark states were proposed by Gell-Mann \cite{1964-Gell-Mann-p214-215} and Zweig \cite{1964-Zweig-p-} at the birth of quark model (QM). In the past fifty years, it has been an extremely intriguing 
research issue of searching for multiquark matter. The light tetraquark $qq\bar q\bar q$ state has been used to 
investigate the scalar mesons below 1 GeV \cite{1977-Jaffe-p281-281}. Since 2003, plenty of charmoniumlike 
states have been observed and the hidden-charm $qc\bar q\bar c$ tetraquark fomalism is extensively 
discussed to explain the nature of these new XYZ states \cite{2016-Chen-p1-121,2011-Chen-p34010-34010,2013-Chen-p45027-45027,2016-Esposito-p1-97,2017-Lebed-p143-194,2017-Guo-p-,2015-Chen-p54002-54002,2017-Chen-p160-160}. 

The doubly hidden-charm/bottom tetraquark $QQ\bar Q\bar Q$ is composed of four heavy quarks. 
Such tetraquark states did not receive much attention in both experimental and theoretical aspects 
 \cite{1975-Iwasaki-p492-492,1981-Chao-p317-317,1982-Ader-p2370-2370,1983-Ballot-p449-451,
1985-Heller-p755-755,2004-Lloyd-p14009-14009,1992-Silvestre-Brac-p2179-2189,1993-Silvestre-Brac-p273-282,
2006-Barnea-p54004-54004,2012-Berezhnoy-p34004-34004}. 
Recently, there are some discussions about the masses and decays of the $QQ\bar Q\bar Q$ states 
\cite{2017-Chen-p247-251,2017-Karliner-p34011-34011,2016-Bai-p-,2016-Wu-p-,2016-Brambilla-p54002-54002,2017-Wang-p432-432,2017-Anwar-p-,2017-Debastiani-p-,2017-Eichten-p-,2017-Hughes-p-}. 
The masses of these $QQ\bar Q\bar Q$ tetraquarks 
are far away from the mass regions of the conventional $Q\bar Q$ mesons and the XYZ states. It will be very easy to distinguish them from the XYZ and $Q\bar Q$ states in the spectroscopy. On the other 
hand, the $QQ\bar Q\bar Q$ states favor the compact tetraquark configuration than the loosely bound hadron 
molecular configuration, since the light mesons can not be exchanged between the two charmonium/bottomonium 
states.
In this paper, we develop a moment QCD sum rule method to calculate the mass spectra for the doubly 
hidden-charm/bottom $cc\bar c\bar c$ and $bb\bar b\bar b$ tetraquark states.

\section{QCD sum rules}
\label{sec-qsr}
In this section we briefly introduce the method of QCD sum rules \cite{1979-Shifman-p385-447,1985-Reinders-p1-1,2000-Colangelo-p1495-1576}. Comparing to the traditional SVZ QCD sum rules, we use another version of QCD sum rules, the moment QCD sum rules in our analyses for the doubly hidden-charm/bottom $QQ\bar Q\bar Q$ tetraquark systems. The moment QCD sum rules have been very successfully used for studying the charmonium and bottomonium mass
spectra \cite{1979-Shifman-p385-447,1979-Shifman-p448-518,1983-Nikolaev-p526-526,1981-Reinders-p109-109,1985-Reinders-p1-1} and determining the heavy quark masses and the strong coupling constant \cite{1997-Jamin-p334-352,2001-Eidemuller-p203-210,2001-Kuhn-p588-602}. 

We start by considering the following two-point correlation functions
\begin{equation}
\begin{split}
\Pi(q)&= i \int d^4xe^{iq\cdot x}\langle 0|T[J(x)J^{\dag}(0)]|0\rangle\, , \\
\Pi_{\mu\nu}(q)&=i\int d^4x e^{iq\cdot x}\langle 0|T [J_\mu(x)J_\nu^\dagger(0)]|0\rangle\, , \\
\Pi_{\mu\nu, \,\rho\sigma}(q)&=i\int d^4x e^{iq\cdot x}\langle 0|T [J_{\mu\nu}(x)J_{\rho\sigma}^\dagger(0)]|0\rangle\, , 
\label{Piq}
\end{split}
\end{equation}
in which the interpolating currents $J(x)$, $J_{\mu}(x)$ and $J_{\mu\nu}(x)$ couple to the scalar, vector and 
tensor states respectively. 

To study the doubly hidden-charm/bottom tetraquarks, we construct the $QQ\bar Q\bar Q$ interpolating currents with four heavy quarks in the compact diquark-antidiquark configuration. 
We use all diquark fields $Q^T_a CQ_b$, $Q^T_a C\gamma_5Q_b$, $Q^T_aC\gamma_\mu\gamma_5Q_b$,
$Q^T_aC\gamma_\mu Q_b$, $Q^T_a C\sigma_{\mu\nu}Q_b$ and $Q^T_aC\sigma_{\mu\nu}\gamma_5Q_b$ 
and consider the Pauli principle to determine the color and flavor structures for the tetraquark operators. Following Refs. \cite{2013-Du-p14003-14003,2017-Chen-p247-251}, we obtain the $QQ\bar Q\bar Q$ tetraquark interpolating currents
as the following. 
The interpolating currents with $J^{PC}=0^{++}$ are
\begin{equation}
\begin{split}
J_1&=Q^T_aC\gamma_5Q_b\bar{Q}_a\gamma_5C\bar{Q}_b^T\, , \\
J_2&=Q^T_aC\gamma_\mu\gamma_5Q_b\bar{Q}_a\gamma^\mu \gamma_5C\bar{Q}_b^T\, , \\
J_3&=Q^T_aC\sigma_{\mu\nu}Q_b\bar{Q}_a\sigma^{\mu\nu}C\bar{Q}^T_b\, , \label{currents1} \\
J_4&=Q^T_aC\gamma_\mu Q_b\bar{Q}_a\gamma^\mu C\bar{Q}_b^T\, , \\
J_5&=Q^T_aCQ_b\bar{Q}_aC\bar{Q}_b^T\, ,
\end{split}
\end{equation}
where $J_1, J_2, J_5$ belong to the symmetric $[\mathbf{6_c}]_{QQ}\otimes[\mathbf{\bar 6_c}]_{\bar Q\bar Q}$ 
color structure while $J_3, J_4$ belong to the antisymmetric 
$[\mathbf{\bar 3_c}]_{QQ}\otimes[\mathbf{3_c}]_{\bar Q\bar Q}$ color structure. 
The interpolating currents with $J^{PC}=0^{-+}$ and $0^{--}$ are
\begin{equation}
\begin{split}
J_1^{\pm}&=Q^T_aCQ_b\bar{Q}_a\gamma_5C\bar{Q}_b^T\pm Q^T_aC\gamma_5Q_b\bar{Q}_aC\bar{Q}_b^T\, , \\
J_2^+&=Q^T_aC\sigma_{\mu\nu}Q_b\bar{Q}_a\sigma^{\mu\nu}\gamma_5C\bar{Q}^T_b\, , \label{currents2}
\end{split}
\end{equation}
in which $J^+_1$ and $J_2^+$ couple to the states with $J^{PC}=0^{-+}$, and $J_1^-$ couples to the states with $J^{PC}=0^{--}$. The currents $J_1^{\pm}$ belong to the symmetric color structure while $J_2^+$ belongs to antisymmetric 
color structure. 
The interpolating currents with $J^{PC}=1^{++}$ and $1^{+-}$ are
\begin{equation}
\begin{split}
J_{1\mu}^{\pm}&=Q^T_aC\gamma_\mu\gamma_5 Q_b\bar{Q}_aC\bar{Q}_b^T
\pm Q^T_aCQ_b\bar{Q}_a\gamma_\mu\gamma_5 C\bar{Q}_b^T\, , \label{currents3}
\\
J_{2\mu}^{\pm}&=Q^T_aC\sigma_{\mu\nu}\gamma_5 Q_b\bar{Q}_a\gamma^\nu C\bar{Q}^T_b
\pm Q^T_aC\gamma^\nu Q_b\bar{Q}_a\sigma_{\mu\nu}\gamma_5C\bar{Q}^T_b\, ,
\end{split}
\end{equation}
in which $J_{1\mu}^{+}$ and $J_{2\mu}^{+}$ couple to the states with $J^{PC}=1^{++}$, and $J_{1\mu}^{-}$ and $J_{2\mu}^{-}$ couple to the states with $J^{PC}=1^{+-}$. The currents $J_{1\mu}^{\pm}$ belong to the symmetric color structure while $J_{2\mu}^{\pm}$ belongs to antisymmetric color structure.  The interpolating currents with $J^{PC}=1^{-+}$ and $1^{--}$ are
\begin{equation}
\begin{split}
J_{1\mu}^{\pm}&=Q^T_aC\gamma_\mu\gamma_5 Q_b\bar{Q}_a\gamma_5C\bar{Q}_b^T
\pm Q^T_aC\gamma_5Q_b\bar{Q}_a\gamma_\mu\gamma_5 C\bar{Q}_b^T \, ,\\
J_{2\mu}^{\pm}&=Q^T_aC\sigma_{\mu\nu}Q_b\bar{Q}_a\gamma^\nu C\bar{Q}^T_b
\pm Q^T_aC\gamma^\nu Q_b\bar{Q}_a\sigma_{\mu\nu}C\bar{Q}^T_b \, ,\label{currents4}
\end{split}
\end{equation}
in which $J_{1\mu}^{+}$ and $J_{2\mu}^{+}$ couple to the states with $J^{PC}=1^{-+}$, and $J_{1\mu}^{-}$ and $J_{2\mu}^{-}$ couple to the states with $J^{PC}=1^{--}$. The currents $J_{1\mu}^{\pm}$ belong to the symmetric color structure while $J_{2\mu}^{\pm}$ belongs to antisymmetric color structure. 
The interpolating currents with $J^{PC}=2^{++}$ are
\begin{equation}
\begin{split}
J_{1\mu\nu}&=Q^T_aC\gamma_\mu Q_b\bar{Q}_a\gamma_\nu C\bar{Q}_b^T
+Q^T_aC\gamma_\nu Q_b\bar{Q}_a\gamma_\mu C\bar{Q}_b^T\, , \label{currents5}
\\
J_{2\mu\nu}&=Q^T_aC\gamma_\mu\gamma_5 Q_b\bar{Q}_a\gamma_\nu\gamma_5 C\bar{Q}_b^T
+Q^T_aC\gamma_\nu\gamma_5Q_b\bar{Q}_a\gamma_\mu\gamma_5 C\bar{Q}_b^T\, ,
\end{split}
\end{equation}
where current $J_{1\mu\nu}$ belongs to the antisymmetric color structure while $J_{2\mu\nu}$ belongs to symmetric color structure. 

At the hadronic level, the correlation functions in Eq.\eqref{Piq} can be described by the dispersion relation 
\begin{align}
\Pi(q^2)=\frac{(q^2)^N}{\pi}\int_{M_H^2}^{\infty}\frac{\mbox{Im}\Pi(s)}{s^N(s-q^2-i\epsilon)}ds+\sum_{n=0}^{N-1}b_n(q^2)^n\, ,
\label{dispersionrelation}
\end{align}
where $M_H$ is the hadron mass and $b_n$ are unknown subtraction constants. A narrow resonance approximation is usually used to describe the spectral function 
\begin{align}
\nonumber
\rho(s)=\frac{1}{\pi}\text{Im}\Pi(s)&=\sum_n\delta(s-m_n^2)\langle0|J|n\rangle\langle
n|J^{\dagger}|0\rangle+\cdots \\
&=f_X^2\delta(s-m_X^2)+\cdots\, , \label{Imaginary}
\end{align}
where ``$\cdots$" represents the excited higher states and continuum contributions and 
$f_X$ is a coupling constant between the interpolating current and hadron state
\begin{equation}
\begin{split}
\langle0|J|X\rangle&=f_X\, ,
\\
\langle0|J_{\mu}|X\rangle&=f_X\epsilon_{\mu}\, ,
\\
\langle0|J_{\mu\nu}|X\rangle&=f_X\epsilon_{\mu\nu}\, ,  \label{coupling parameters}
\end{split}
\end{equation}
in which $\epsilon_{\mu}$ and $\epsilon_{\mu\nu}$ are the polarization vector and tensor, respectively. 
To pick out the contribution of the lowest lying resonance in Eq. \eqref{Imaginary}, we define moments  
in Euclidean region $Q^2=-q^2>0$ \cite{1985-Reinders-p1-1,2014-Chen-p201-215}:
\begin{align}
M_n(Q^2_0)=\frac{1}{n!}\bigg(-\frac{d}{dQ^2}\bigg)^n\Pi(Q^2)|_{Q^2=Q_0^2}
&=\int_{16m_Q^2}^{\infty}\frac{\rho(s)}{(s+Q^2_0)^{n+1}}ds\,  \label{moment}
\\
&=\frac{f_X^2}{(m_X^2+Q_0^2)^{n+1}}\big[1+\delta_n(Q_0^2)\big]\,, \label{Phemoment}
\end{align}
in which $\delta_n(Q_0^2)$ contains the contributions of higher states and continuum.
It tends to zero as $n$ goes to infinity. We consider the following ratio to eliminate $f_X$ 
in Eq.~\eqref{Phemoment}
\begin{align}
r(n,Q_0^2)\equiv\frac{M_{n}(Q_0^2)}{M_{n+1}(Q_0^2)}=\big(m_X^2+Q_0^2\big)
\frac{1+\delta_{n}(Q_0^2)}{1+\delta_{n+1}(Q_0^2)}. \label{ratio}
\end{align}
One expects $\delta_{n}(Q_0^2)\cong\delta_{n+1}(Q_0^2)$ for sufficiently large $n$ 
to suppress the contributions of higher states and continuum \cite{1985-Reinders-p1-1}. 
Then hadron mass of the lowest lying resonance $m_X$ is then extracted as 
\begin{align}
m_X=\sqrt{r(n,Q_0^2)-Q_0^2}\, . \label{mass}
\end{align}

Using the operator production expansion (OPE) method, the two-point function can also be evaluated 
at the quark-gluonic level as a function of various QCD parameters. In the fully heavy tetraquark 
systems, we only need to calculate the perturbative term and the gluon condensate contributions 
to the correlation functions. One can find the results of the moments $M_n(Q^2_0)$ in 
Ref. \cite{2017-Chen-p247-251}.

\section{Numerical results}
We perform the numerical analyses by using the following values of parameters \cite{2014-Olive-p90001-90001,1996-Narison-p162-172,2003-Ioffe-p229-241,2010-Narison-p559-559}
\begin{equation}
\begin{split}
 m_c(\overline{\rm MS})&=1.27\pm0.03~\text{GeV}, \,
\\
\label{parameters} 
m_b(\overline{\rm MS})&=4.18\pm0.03~\text{GeV}\, ,
\\
 \langle g_s^2GG\rangle&=(0.48\pm0.14)~\text{GeV}^4\, .
\end{split}
\end{equation}
To provide reliable moment sum rule analyses, one needs to find suitable working regions of the two parameters
$n$ and $Q_0^2$ in the ratio $r(n,Q_0^2)$. We define $\xi=Q^2_0/16m_c^2$ for $cc\bar c\bar c$ and 
$Q^2_0/m_b^2$ for $bb\bar b\bar b$ systems for convenience. These two parameters will affect the pole 
contribution and the OPE convergence. For small value of $\xi$, the high dimension condensates 
in OPE will give large contributions, and thus leading to bad OPE convergence \cite{1981-Reinders-p109-109,
1985-Reinders-p1-1}. However, a larger value of $\xi$ means slower convergence of $\delta_{n}(Q_0^2)$ 
in Eq. \eqref{Phemoment}. Such behavior can be compensated by $n$: the OPE convergence becomes 
good for small $n$ while $\delta_{n}(Q_0^2)$ tends to zero for sufficiently large $n$. 
One needs to find suitable working regions for $(n, \xi)$ where the lowest lying resonance dominates the 
moments and the OPE converges well.

\begin{figure}[hbt]
\begin{center}
\scalebox{0.65}{\includegraphics{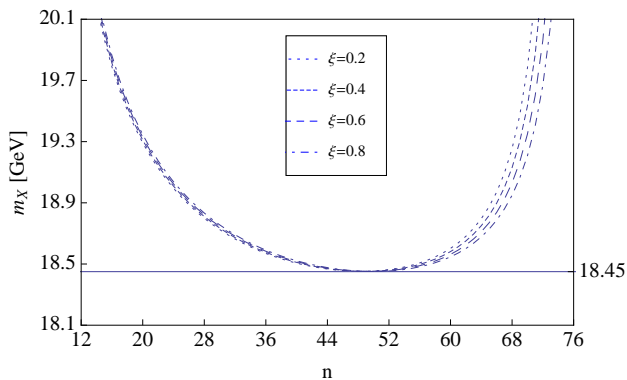}}
\caption{Hadron mass $m_{X_b}$ for $J_{1}(bb\bar b\bar b)$ with $J^{PC}=0^{++}$, as a function of $n$
for different value of $\xi$.} \label{figbbbb0++}
\end{center}
\end{figure}

As an example, we use the interpolating current $J_1$ with $J^{PC}=0^{++}$ in Eq. \eqref{currents1} to perform 
numerical analyses. Requiring the perturbative term to be larger than the gluon condensate term, we obtain 
upper limits $n_{max}$, which increases with respect to the value of $\xi$.
We show the hadron mass $m_{X_b}$ as a function of $n$ for $\xi=0.2, 0.4, 0.6, 0.8$ in Fig.~\ref{figbbbb0++}.
One notes that the mass curves have plateaus which provide stable mass prediction
\begin{equation}
m_{X_b}=(18.45 \pm 0.15) \,\mbox{GeV}\, ,
\end{equation}
in which the error comes from the uncertainties of $\xi$, the heavy quark mass and
the gluon condensate in Eq.~\eqref{parameters}. Using the interpolating currents in Eqs.~\eqref{currents1}--\eqref{currents5}, we perform numerical analyses for all $cc\bar c\bar c$ and $bb\bar b\bar b$ systems 
with various quantum numbers. We collect the numerical results in Table \ref{tablemass}. 
It is shown that the negative parity states ($J^{PC}=0^{-+}, 0^{--}, 1^{-+}, 1^{--}$) are slightly heavier than 
the positive parity states ($J^{PC}=0^{++}, 1^{++}, 1^{+-}, 2^{++}$). 
\begin{table}
\begin{center}
\begin{tabular*}{8.4cm}{cccc}
\hline
~~~~$J^{PC}$ ~~~~& Currents &~~~~ $m_{X_{c}}$\mbox{(GeV)} ~~~~&~~~~ $m_{X_{b}}$\mbox{(GeV)} ~~~~ \\
\hline
$0^{++}$      & $J_1$               &  $6.44\pm0.15$                  & $18.45\pm0.15$  \\
              & $J_2$               &  $6.59\pm0.17$                & $18.59\pm0.17$        \\
                & $J_3$               &  $6.47\pm0.16$                &$18.49\pm0.16$     \\
                    & $J_4$               &  $6.46\pm0.16$                 &$18.46\pm0.14$       \\
                       & $J_5$               &  $6.82\pm0.18$                  & $19.64\pm0.14$
\vspace{4pt}\\
$0^{-+}$        & $J_1^+$            &  $6.84\pm0.18$                 & $18.77\pm0.18$ \\
                     & $J_2^+$            &  $6.85\pm0.18$                 & $18.79\pm0.18$
\vspace{4pt}\\
$0^{--}$       & $J_1^-$              &  $6.84\pm0.18$                & $18.77\pm0.18$
\vspace{4pt}\\
$1^{++}$      & $J_{1\mu}^+$    & $6.40\pm0.19$                & $18.33\pm0.17$  \\
                    & $J_{2\mu}^+$    & $6.34\pm0.19$                & $18.32\pm0.18$
\vspace{4pt}\\
$1^{+-}$       & $J_{1\mu}^-$     & $6.37\pm0.18$                & $18.32\pm0.17$  \\
                    & $J_{2\mu}^+$    & $6.51\pm0.15$                & $18.54\pm0.15$
\vspace{4pt}\\
$1^{-+}$       & $J_{1\mu}^+$    & $6.84\pm0.18$                 & $18.80\pm0.18$ \\
                    & $J_{2\mu}^+$    & $6.88\pm0.18$               & $18.83\pm0.18$
\vspace{4pt}\\
$1^{--}$        & $J_{1\mu}^-$     & $6.84\pm0.18$                & $18.77\pm0.18$  \\
                    & $J_{2\mu}^-$     & $6.83\pm0.18$              & $18.77\pm0.16$
\vspace{4pt}\\
$2^{++}$       & $J_{1\mu\nu}$  & $6.51\pm0.15$                & $18.53\pm0.15$  \\
                     & $J_{2\mu\nu}$  & $6.37\pm0.19$                & $18.32\pm0.17$  \\
\hline
\end{tabular*}
\caption{Mass spectra for the $cc\bar c\bar c$ and $bb\bar b\bar b$
tetraquarks. \label{tablemass}}
\end{center}
\end{table}

\begin{figure}[hbt]
\begin{center}
\scalebox{0.5}{\includegraphics{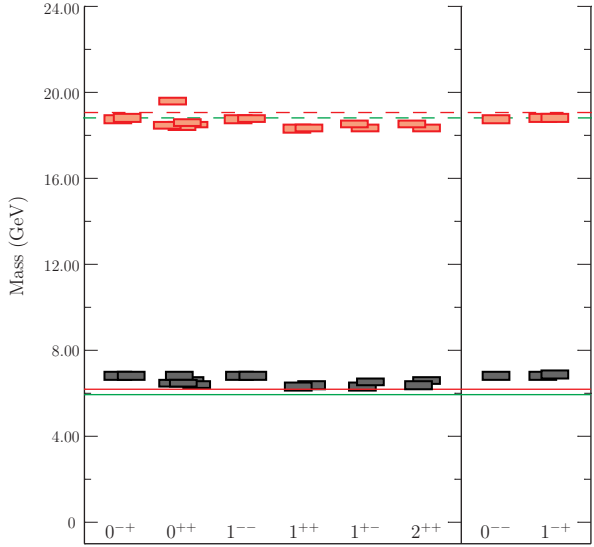}}
\end{center}
\caption{Summary of the doubly hidden-charm/bottom tetraquark spectra labelled by $J^{PC}$. 
The green and red solid (dashed) lines indicate the $\eta_c(1S)\eta_c(1S)$ ($\eta_b(1S)\eta_b(1S)$) and $J/\psi J/\psi$
($\Upsilon(1S)\Upsilon(1S)$) thresholds, respectively.} \label{fig:spectra}
\end{figure}

It is interesting to compare the mass spectra with the corresponding two-meson mass thresholds. 
As shown in Fig.~\ref{fig:spectra}, the masses of $bb\bar b\bar b$ tetraquarks are below the $\eta_b(1S)\eta_b(1S)$ 
threshold while all $cc\bar c\bar c$ tetraquarks lie above the $\eta_c(1S)\eta_c(1S)$ threshold. 
The two bottomonium mesons decays for the $bb\bar b\bar b$ tetraquarks are thus forbidden by the kinematics.
For the doubly hidden-charm $cc\bar c\bar c$ tetraquarks, they can decay via the
spontaneous dissociation mechanism by considering the restrictions of the symmetries. 
In Table \ref{ccccdecay}, we collect the possible $S$-wave and $P$-wave dissociation decay channels for the $cc\bar c\bar c$ states.

In principle, the $bb\bar b\bar b$ tetraquark can also decay into $B^{(\ast)}\bar B^{(\ast)}$ via a heavy quark pair annihilation and a light quark pair creation processes. The suppression by the annihilation of a heavy quark pair 
will be compensated by the large phase space factor. Such $B^{(\ast)}\bar B^{(\ast)}$ decay modes may dominate 
the total width of the doubly hidden-bottom $bb\bar b\bar b$ tetraquark state.

\begin{table}[hbt]
\begin{center}
\begin{tabular*}{9.2cm}{ccc}
\hline
$J^{PC}$       & S-wave              & P-wave   \\
\hline
$0^{++}$  & $\eta_c(1S)\eta_c(1S)$, $ J/\psi J/\psi$ &  $\eta_c(1S)\chi_{c1}(1P)$, $ J/\psi h_c(1P)$
\vspace{6pt}\\
$0^{-+}$        & $\eta_c(1S)\chi_{c0}(1P)$, $ J/\psi  h_c(1P)$    &  $ J/\psi J/\psi$
\vspace{6pt}\\
$0^{--}$         & $ J/\psi\chi_{c1}(1P)$    &  $ J/\psi\eta_c(1S)$
\vspace{6pt}\\
$1^{++}$       &  $-$   &  $ J/\psi h_c(1P)$,  $\eta_c(1S)\chi_{c1}(1P)$,  \\
                     &                                                 &  $\eta_c(1S)\chi_{c0}(1P)$
\vspace{6pt}\\
$1^{+-}$        & $ J/\psi\eta_c(1S)$      &  $ J/\psi\chi_{c0}(1P)$, $ J/\psi\chi_{c1}(1P)$,  \\
                     &                                                &  $\eta_c(1S) h_c(1P)$
\vspace{6pt}\\
$1^{-+}$  & $ J/\psi h_c(1P)$, $\eta_c(1S)\chi_{c1}(1P)$ &   $-$
\vspace{6pt}\\
$1^{--}$         & $ J/\psi\chi_{c0}(1P)$, $ J/\psi\chi_{c1}(1P)$,     &  $ J/\psi\eta_c(1S)$ \\
                     & $\eta_c(1S) h_c(1P)$                                                                &
\\ \hline
\end{tabular*}
\caption{Possible decay modes of the $cc\bar c\bar c$ states by spontaneous dissociation
into two charmonium mesons. \label{ccccdecay}}
\end{center}
\end{table}

\section{Summary}
In this paper, we have calculated the mass spectra for the doubly hidden-charm/bottom $cc\bar c\bar c$ 
and $bb\bar b\bar b$ tetraquark states by using the moment QCD sum rule method. 
Our results show that the $cc\bar c\bar c$ tetraquarks lie above the two charmonium spontaneous dissociation 
thresholds and thus can mainly decay into two charmonium mesons. We suggest to search for these doubly 
hidden-charm $cc\bar c\bar c$ states in the $J/\psi J/\psi$ and $\eta_c(1S)\eta_c(1S)$ channels. 
For the $bb\bar b\bar b$ tetraquarks, their masses are lower than the $\eta_b(1S)\eta_b(1S)$ threshold 
so that the two bottomonium mesons decays are kinematical forbidden. These $bb\bar b\bar b$ tetraquark, 
if exist, may be very narrow and stable. In the near future, these doubly hidden-charm/bottom $cc\bar c\bar c$ 
and $bb\bar b\bar b$ tetraquark states can be searched for at facilities such as LHCb, CMS, RHIC and the
forthcoming BelleII.

\section*{Acknowledgments}
This project is supported by the Chinese National Youth Thousand Talents Program; the Natural Sciences and Engineering Research Council of Canada (NSERC);  the National Natural Science
Foundation of China under Grants No. 11475015, 373
No. 11375024, No. 11222547, No. 11175073, 374 No. 11575008, and No. 11621131001; the 973 program; 375 the Ministry of Education of China (SRFDP under Grant 376 No. 20120211110002 and the Fundamental Research 377 Funds for the Central Universities); and the National 378 Program for Support of Top-Notch Youth Professionals.


\end{document}